\documentclass[aps,showpacs,prl,reprint,superscriptaddress,longbibliography]{revtex4-1}

\usepackage[colorlinks=true,linkcolor=blue,citecolor=blue,urlcolor=blue]{hyperref}

\usepackage{setspace} 
\usepackage{graphicx}
\usepackage{amsmath}
\usepackage{color}
\usepackage{amsmath}
\usepackage{amssymb}
\usepackage{verbatim}
\usepackage{latexsym}
\usepackage{enumerate} 
\usepackage{bm} 
\usepackage[caption=false]{subfig}
\begin{document}

\title{Temperature Dependence of the Energy Levels of Methylammonium Lead Iodide Perovskite from First Principles}

\author{Wissam A.\ Saidi} \email{alsaidi@pitt.edu}
\affiliation{Department of Mechanical Engineering and Materials Science,
  University of Pittsburgh, Pittsburgh 15261, USA}

\author{Samuel Ponc\'{e}} 
\affiliation{Department of Materials, University of Oxford, Parks Road,
  Oxford OX1 3PH, United Kingdom}

\author{Bartomeu Monserrat} \affiliation{Department of Physics and Astronomy, Rutgers University, Piscataway, New Jersey 08854-8019, USA} \affiliation{TCM Group, Cavendish Laboratory, University of Cambridge, J.\ J.\ Thomson Avenue, Cambridge CB3 0HE, United Kingdom}

\date{\today}


\begin{abstract}
Environmental effects and intrinsic energy-loss processes lead to fluctuations in the operational temperature of solar cells, which can profoundly
influence their power conversion efficiency. Here we determine from first principles the 
effects of temperature on the band gap and band edges of the hybrid pervoskite 
CH$_3$NH$_3$PbI$_3$ by accounting
for electron-phonon coupling and thermal expansion. From $290$ to
$380$~K, the computed band gap change of $40$~meV coincides with the
experimental change of $30$-$40$~meV. The calculation of electron-phonon coupling in 
CH$_3$NH$_3$PbI$_3$ is particularly intricate, as the commonly used Allen-Heine-Cardona 
theory overestimates the band gap change with temperature, and 
excellent agreement with experiment is only obtained when including high-order terms in
the electron-phonon interaction. We also find that spin-orbit coupling
enhances the electron-phonon coupling strength, but that the inclusion of
nonlocal correlations using hybrid functionals has little effect. We reach similar
conclusions in the metal-halide perovskite CsPbI$_3$.
Our results unambiguously confirm for the first time the importance of high-order terms 
in the electron-phonon coupling by direct comparison with experiment.
\end{abstract}

\maketitle


Metal-halide hybrid perovskites are promising candidates as light absorbers for the next generation of solid state hybrid photovoltaic devices~\cite{Kojima_JACS_2009,Kim_2012,Lee02112012,Burschka_2013,snaith_2013,Xing_2013,Hodes18102013}. Their solar power conversion efficiency has reached $22.1$\%~\cite{cell_eff}, an impressive achievement considering that the first perovskite-based solar cell was only made seven years ago~\cite{Kojima_JACS_2009}. These light-absorbing materials belong to the perovskite family ABX$_3$, where A is a cation located at the center of a cube formed from corner-sharing BX$_6$ octahedra.

Methylammonium lead iodide MAPbI$_3$, where MA represents the organic cation CH$_3$NH$_3^{+}$, has an optimal band gap for light absorption of $1.6$~eV and exhibits large electron and hole diffusion lengths, which make it a prime candidate for solar cell applications~\cite{Lee02112012,Burschka_2013}. 
Another factor determining the efficiency of a solar cell is the band alignment between the light absorber and its transport layers~\cite{Marcus_1965}.
As an example, a common electron transport layer is TiO$_2$, which has a conduction band that is only $0.1$~eV lower than that of MAPbI$_3$~\cite{tio2_mapbi3_band_alignment}.

Solar cell devices are subject to significant temperature variations
under normal operating conditions~\cite{SZE_NG}. Ambient temperatures
can oscillate between $250$ and $310$~K, and energy-loss mechanisms
associated with the conversion of solar power to electric power can
increase the operating temperature of the cell above
$340$~K~\cite{SZE_NG,Energy_conference,solar_cell_operating_temperature}.
These temperature fluctuations can affect both the absorption onset
and band alignment, and therefore the performance of the solar cell,
as was demonstrated for MAPbI$_3$-based solar cells, which showed an
open circuit voltage drop from $1.01$~V to $0.83$~V as temperature increased from 300 to 360~K~\cite{Zhang_2015}.
Recently, two  experimental studies have
investigated the temperature-dependent properties of MAPbI$_3$ and
reported changes in the band gap reaching $40$~meV in the temperature
range of solar cell
operation~\cite{Foley_perovskites_2015,Milot_2015_temperature_mapbi3},
and even larger changes exceeding $100$~meV in the individual band
edges~\cite{Foley_perovskites_2015}.

While the properties of MAPbI$_3$ are well-characterized both
theoretically and experimentally, there is an increasing research
effort to find alternatives to this light
absorber~\cite{mapbi3_gap_tuning_experiment,dft_perovskites,filip_chemical_substitution,lead_free_perovskites}.
with the aim of tuning the physical properties of the solar cells and 
moving towards lead-free devices. In this context, it would be desirable
to accurately calculate the effects of temperature changes  
to aid in the computational design of novel perovskite solar cell
materials under real working conditions. Unfortunately, the organic-inorganic 
nature of MAPbI$_3$, together with the presence of the heavy Pb atoms, make
it a challenging system to study from first
principles~\cite{dft_tio2_mapbi3,dft_tio2_mapbi3_ii,experiment_mapbi3_gap,mapbi3_dft_nsoc_gap,mapbi3_dft_soc_gap,mapbi3_gw_soc_gap,mapbi3_scgw_soc_gap,mapbi3_gw_soc_gap_giustino}.

In this Letter, we determine the temperature evolution of the band gap and 
band edges in MAPbI$_3$ using first-principles methods while accounting 
for both electron-phonon coupling and thermal expansion. Our
results accurately reproduce the band gap opening observed experimentally
in MAPbI$_3$ upon increasing
temperature~\cite{Foley_perovskites_2015,Milot_2015_temperature_mapbi3}.
We further show that the commonly used Allen-Heine-Cardona (AHC)
theory~\cite{ahc_1976,Allen1981}, which successfully captures the
low-order terms in electron-phonon coupling and can be used to explain the
temperature dependence of band gaps in multiple systems, fails in MAPbI$_3$
because high-order terms in the electron-phonon interaction make significant
contributions. Furthermore, we show that spin-orbit coupling (SOC) must be 
included to
accurately describe the electron-phonon coupling strength of this
system, while nonlocal correlation effects incorporated using a
hybrid functional do not yield significant changes.  Our final results
are in excellent agreement with experimental data, thus verifying that
thermal effects can be reliably computed using
first-principles methods to design novel perovskite solar cells.
Additionally, our demonstration that a low-order treatment of the electron-phonon interactions fails for the band edges of perovskite solar cell materials raises the question of whether this would also be true for other electronic states in the Brillouin zone. Quantities such as electronic transport depend on the electron-phonon interaction averaged over the entire Brillouin zone, and therefore it would be interesting to study whether the standard approach to these calculations based on a low-order treatment of electron-phonon coupling is valid.

{\it Methodology.} Our calculations are based on density functional
theory (DFT)~\cite{hohenberg_and_kohn,kohn_and_sham,dft_rev_mod_phys}
using the {\sc vasp}~\cite{vasp1,vasp2,vasp3,vasp4} and {\sc
  abinit}~\cite{Gonze2016} software packages. We use the semi-local
Perdew-Burke-Ernzerhof (PBE)~\cite{PBE} exchange-correlation functional, 
as well as the hybrid Heyd-Scuseria-Ernzerhof (HSE)~\cite{hse06_functional,hse06_functional_erratum} functional,
with and without SOC. The computational details are provided in the
Supporting Information. To make our calculations tractable, we focus
on the high temperature cubic phase of MAPbI$_3$, but we expect that our 
results are also applicable to the tetragonal phase that is stable below 
$330$~K~\cite{mapbi3_phase_transitions_i,experiment_mapbi3_gap,mapbi3_phase_transitions_ii}, as experimental results show that the band gap evolution with 
temperature transitions smoothly between the cubic and tetragonal
phases~\cite{Foley_perovskites_2015,Milot_2015_temperature_mapbi3}.
The cubic phase exhibits some unstable modes, and we have set their amplitudes
to zero in the calculations reported. These modes can contribute to the 
strength of electron-phonon coupling~\cite{walsh_anharmonic_elph_coupling_arxiv},
but as they represent only a small fraction of all modes, we expect their contribution to the temperature dependence of the band gap to be small. Calculations in which these modes are allowed to have nonzero amplitudes indicate that the neglect of the unstable modes leads to an error that is smaller than the statistical uncertainty in our calculations (see Supporting Information). 

{\it Allen-Heine-Cardona theory.}
The coupling between electronic states and nuclear vibrations
due to quantum zero-point (ZP) motion and thermal motion renormalizes
the electronic eigenenergies $\epsilon_{\mathbf{k}n}$ as
\begin{equation}
\epsilon_{\mathbf{k}n}(T)=\frac{1}{\mathcal{Z}}\sum_{\mathbf{s}}\langle\chi_{\mathbf{s}}|\epsilon_{\mathbf{k}n}|\chi_{\mathbf{s}}\rangle e^{-E_{\mathbf{s}}/k_{\mathrm{B}}T}, \label{eq:epsilon_temperature}
\end{equation}
where $|\chi_{\mathbf{s}}\rangle$ is the vibrational eigenstate $\mathbf{s}$ 
with eigenvalue $E_{\mathbf{s}}$, $\mathcal{Z}=\sum_{\mathbf{s}}e^{-E_{\mathbf{s}}/k_{\mathrm{B}}T}$ is the partition function, $T$ is the temperature, and $k_{\mathrm{B}}$ is Boltzmann constant. 
The evaluation of Eq.~(\ref{eq:epsilon_temperature}) is challenging,
and has only recently become fully amenable to first-principles methods~\cite{giustino_2010,cannuccia_2011,patrick_2013,monserrat_elph_diamond_silicon_2014,ponce_abinit_vs_yambo_2014,giustino_elph_review_arxiv}. The vast majority of these calculations rely on a low-order expansion of the electronic energy as a function of the atomic displacements, leading to a particularly convenient expression for the change in the electronic energy $\Delta \epsilon_{\mathbf{k}n}(T)\triangleq\epsilon_{\mathbf{k}n}(T) -\epsilon_{\mathbf{k}n}(0)$ at temperature $T$:
\begin{equation}
\Delta \epsilon_{\mathbf{k}n}(T) = \frac{1}{N_{\mathbf{q}}}\sum_{\mathbf{q},\nu}\frac{a^{(2)}_{\mathbf{q}\nu;\mathbf{q}\nu}}{\omega_{\mathbf{q}\nu}}\left[\frac{1}{2}+n_{\mathrm{B}}(\omega_{\mathbf{q}\nu},T)\right], \label{eq:ahc}
\end{equation}
where $\mathbf{q}$ and $\nu$ are the phonon quantum numbers, $N_{\mathbf{q}}$ is the number of points in the vibrational Brillouin zone (BZ), $a^{(2)}_{\mathbf{q}\nu;\mathbf{q}\nu}$ is the second-order electron-phonon coupling constant,
$\omega_{\mathbf{q}\nu}$ is the harmonic frequency, and $n_{\mathrm{B}}$ is a Bose-Einstein factor. The coupling constants can be evaluated using finite displacements (FD) in conjunction with supercells~\cite{phonon_finite_displacement,non_diagonal}, or using density functional perturbation theory (DFPT) by invoking the rigid-ion approximation,~\footnote{A description of the use of the rigid-ion approximation in the context of the AHC theory can be found in Ref.~\onlinecite{Ponce2014a}} which leads to the so-called AHC theory~\cite{ahc_1976,Allen1981}. Further details of these methods are provided in the Supporting Information.

The approximation in Eq.~(\ref{eq:ahc}) is computationally tractable 
as it allows a fine sampling of the vibrational BZ because phonon
modes can be treated independently. The AHC theory within DFPT can be
used to sample the vibrational BZ using very fine $\mathbf{q}$-point
grids, readily obtaining converged results~\cite{ponce_2015}. The
recent introduction of nondiagonal supercells has allowed calculations
using the FD method to reach levels of convergence approaching those
of DFPT~\cite{non_diagonal}. In the Supporting Information we present a
comparison of both approaches, which deliver similar results.
We therefore infer that it is safe to apply the rigid-ion approximation in MAPbI$_3$.

\begin{figure} \centering
\includegraphics[scale=0.43]{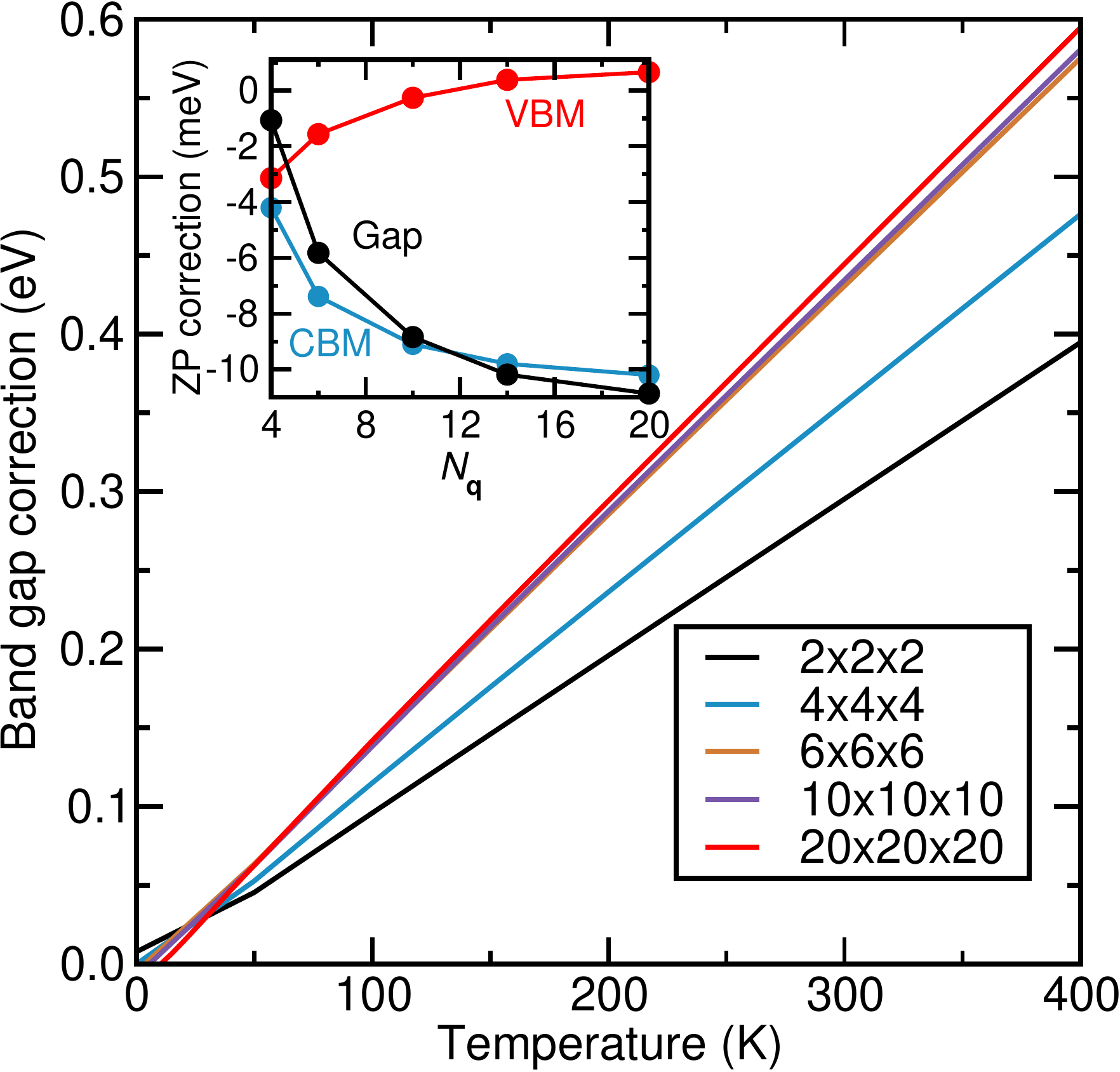}
\caption{Temperature dependence of the band gap of MAPbI$_3$ evaluated
  using the AHC approach with different BZ grids of sizes
  $N_{\mathbf{q}}^3$. The inset shows the convergence of the VBM, CBM
  and band gap ZP corrections at the {\bf R}-point as a function of the
  vibrational BZ grid linear size.}
\label{fig:abinit_convergence}
\end{figure}

{\it Results and Discussion.}  We apply the AHC theory using DFPT in
conjunction with the PBE functional~\cite{PBE} to compute the ZP and
finite-temperature renormalization of the direct minimum gap of
MAPbI$_3$, located at the BZ boundary point
$\mathbf{R}$. Figure~\ref{fig:abinit_convergence} shows
the temperature dependence of the band gap change obtained using
different $\mathbf{q}$-point grids to sample the vibrational BZ, and
the inset shows the corresponding ZP corrections arising from quantum
motion of the valence band maximum (VBM), conduction band minimum (CBM) and band gap.
 Appreciable convergence is reached on the temperature range
0-400~K with a $6\times6\times6$ \textbf{q}-point grid. However, as
seen in the inset of Fig.~\ref{fig:abinit_convergence}, the very small
ZP correction needs a denser \textbf{q}-point grid of
$20\times20\times20$ to be accurately converged in the meV range.  

Figure~\ref{fig:abinit_convergence} shows that the electron-phonon
coupling widens the band gap with increasing temperature, which is
opposite to the behavior of most materials where temperature 
reduces the size of the band gap~\cite{Cardona_review_2005}.
We find that thermal expansion further widens the band gap by $50$~meV as temperature
increases from $0$ to $400$~K, in agreement
with earlier first-principles calculations~\cite{Foley_perovskites_2015,WAS_2016}.
The band gap  opening is in qualitative agreement with two recent experiments 
which also observe gap opening with increasing
temperature~\cite{Foley_perovskites_2015,Milot_2015_temperature_mapbi3}.
However, our calculated band gap change severely overestimates
the one observed experimentally: calculations result in a band gap change
of $120$~meV between $290$~K and $380$~K, while the
experimentally observed increase is only 
$30$--$40$~meV~\cite{Foley_perovskites_2015,Milot_2015_temperature_mapbi3}.
Figure S5 in the Supporting Information shows the temperature dependence of 
the band gaps at other points in the BZ, all indicate that the large 
change with temperature is not limited to the {\bf R} zone boundary point, 
but rather is a general feature of the AHC theory in MAPbI$_3$. 

To better understand the large discrepancy between experiment and
theory, we examine temperature effects on the individual band edges at
the {\bf R} point. Ignoring thermal expansion, in the temperature range from $290$~K 
to $380$~K, the VBM decreases by $104$~meV while the CBM increases by $43$~meV.
Experimentally, the VBM is observed to decrease by $110$~meV, in agreement 
with our calculations, but at variance with our calculations, the CBM also
shifts down by $77$~meV~\cite{Foley_perovskites_2015}.


\begin{figure} \centering
\includegraphics[scale=0.43]{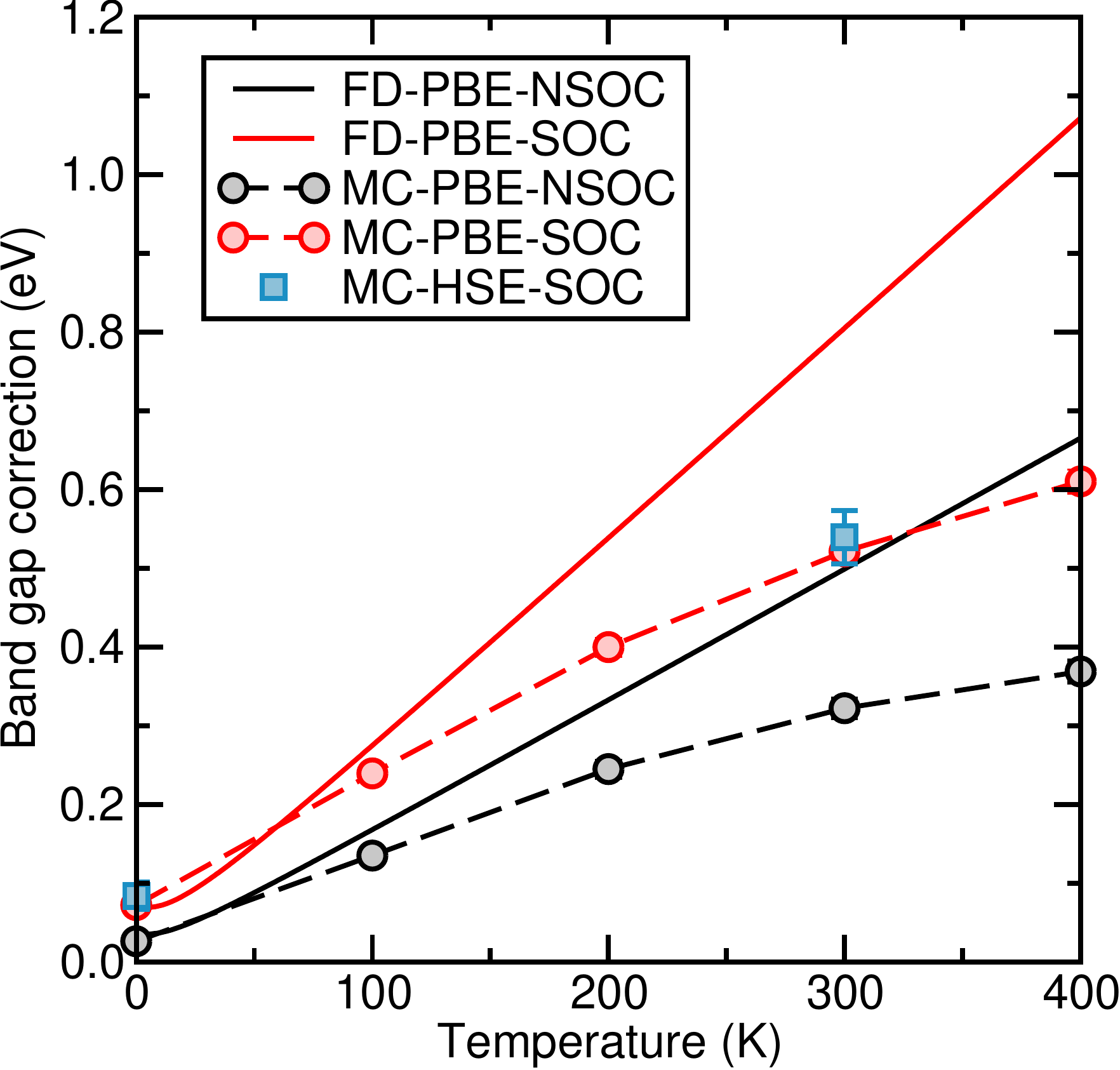}
\caption{Temperature dependence of the band gap of MAPbI$_3$ evaluated
  using the PBE functional and the FD approach without (NSOC) and with (SOC) 
  spin-orbit coupling, and using the MC approach without and with SOC. Additional MC calculations are performed with the HSE functional and including SOC. The 
 calculations are 
  performed for the primitive cell, equivalent to sampling the $\boldsymbol{\Gamma}$-point 
  only. The statistical error bars in the MC results are included in all data points, but their size is smaller than the symbol size for some.}
\label{fig:method}
\end{figure}

The chemical character of the CBM is dominated by Pb atoms in
MAPbI$_3$, and this suggests that the disagreement between theory and
experiment might be due to the neglect of relativistic effects in our
calculations so far. Indeed, SOC is needed to accurately calculate the
static lattice
band gap~\cite{mapbi3_dft_soc_gap,mapbi3_gw_soc_gap_giustino}. To test
the importance of SOC on our results, we have performed
electron-phonon coupling calculations using the FD
approach. Figure~\ref{fig:method} compares the temperature dependence
of the MAPbI$_3$ band gap without (FD-PBE-NSOC) and with (FD-PBE-SOC)
spin-orbit coupling. In these calculations, we used the primitive cell
(equivalent to sampling only the $\boldsymbol{\Gamma}$-point of the
vibrational BZ), and thus are not converged. Nevertheless, these
results clearly demonstrate that SOC {\it enhances} the strength of
electron-phonon coupling, almost doubling the band gap
change with temperature. This shows that, although the proper description
of electron-phonon coupling in MAPbI$_3$ requires SOC, its inclusion
{\it worsens} the agreement between AHC theory and experiment.

Another possible explanation for the discrepancy between the AHC theory
and experiment is electron correlation, which can modify the strength
of electron-phonon coupling in gapped systems~\cite{antonius_gw_elph,gw_thermal_lines}.
To test this, we have performed additional calculations using the hybrid HSE
functional~\cite{hse06_functional,hse06_functional_erratum}. Our
results show that the inclusion of nonlocal correlation does
not significantly affect the electron-phonon coupling strength. 
As a consequence, in the rest of the paper all calculations are 
performed using PBE+SOC.


Our analysis suggests that neither the inclusion of SOC nor a better
description of exchange-correlation effects in the DFT functional, nor
the rigid-ion approximation are
responsible for the discrepancy between the AHC theory and experiment. The
only possibility left is that the commonly used low-order expansion of
electron-phonon coupling fails in MAPbI$_3$. Despite the successes of
the AHC theory to describe the temperature dependence of the band
structures of a wide range of materials~\cite{giustino_2010,elph_Si_nano,
monserrat_elph_diamond_silicon_2014,ponce_2015}, recent calculations question
the validity of the AHC theory in helium at terapascal
pressures~\cite{helium}, molecular crystals at ambient
conditions~\cite{molec_crystals_elph}, in the perovskite CsSnI$_3$~\cite{patrick_cssni3_2015}, and even (albeit moderately) in crystals like diamond~\cite{Antonius2015}.

To assess the importance of high-order terms in the evaluation of
electron-phonon coupling in MAPbI$_3$, we compute
Eq.~(\ref{eq:epsilon_temperature}) directly using a Monte Carlo (MC)
method~\cite{patrick_2013,helium}. We show the results in Fig.~\ref{fig:method},
obtained by sampling the vibrational BZ at the $\boldsymbol{\Gamma}$-point only. 
The ZP quantum corrections to the band gaps are similar between the 
FD and MC approaches, but their temperature dependence differs 
substantially. For example, for the curves including SOC, at
$400$~K, the FD change in the band gap overestimates the MC change by
$460$~meV. Thus, in MAPbI$_3$ high-order terms in the
electron-phonon coupling are important, and the commonly used
Eq.~(\ref{eq:ahc}), with either DFPT or FD, fails in this
system. Interestingly, to the authors knowledge, this is the first 
instance of the failure of the AHC theory that can be verified by
direct comparison to experimental data.

The MC method is a real space method, and the convergence of the
results with system size (equivalent to the convergence with respect
to the BZ grid size) is particularly difficult. The computation of the
electron-phonon coupling with cross-terms requires large diagonal
supercells with sizes limited by the
maximum number of atoms that can be realistically included in a DFT
calculation. For MAPbI$_3$, the band gap correction in the temperature
range 290--380~K is sufficiently converged with $3\times3\times3$
primitive cells, containing $324$ atoms. Finite size effects are
discussed in the Supporting Information based on the convergence of the
AHC method in Fig.~\ref{fig:abinit_convergence}.

\begin{figure} \centering
\includegraphics[scale=0.43]{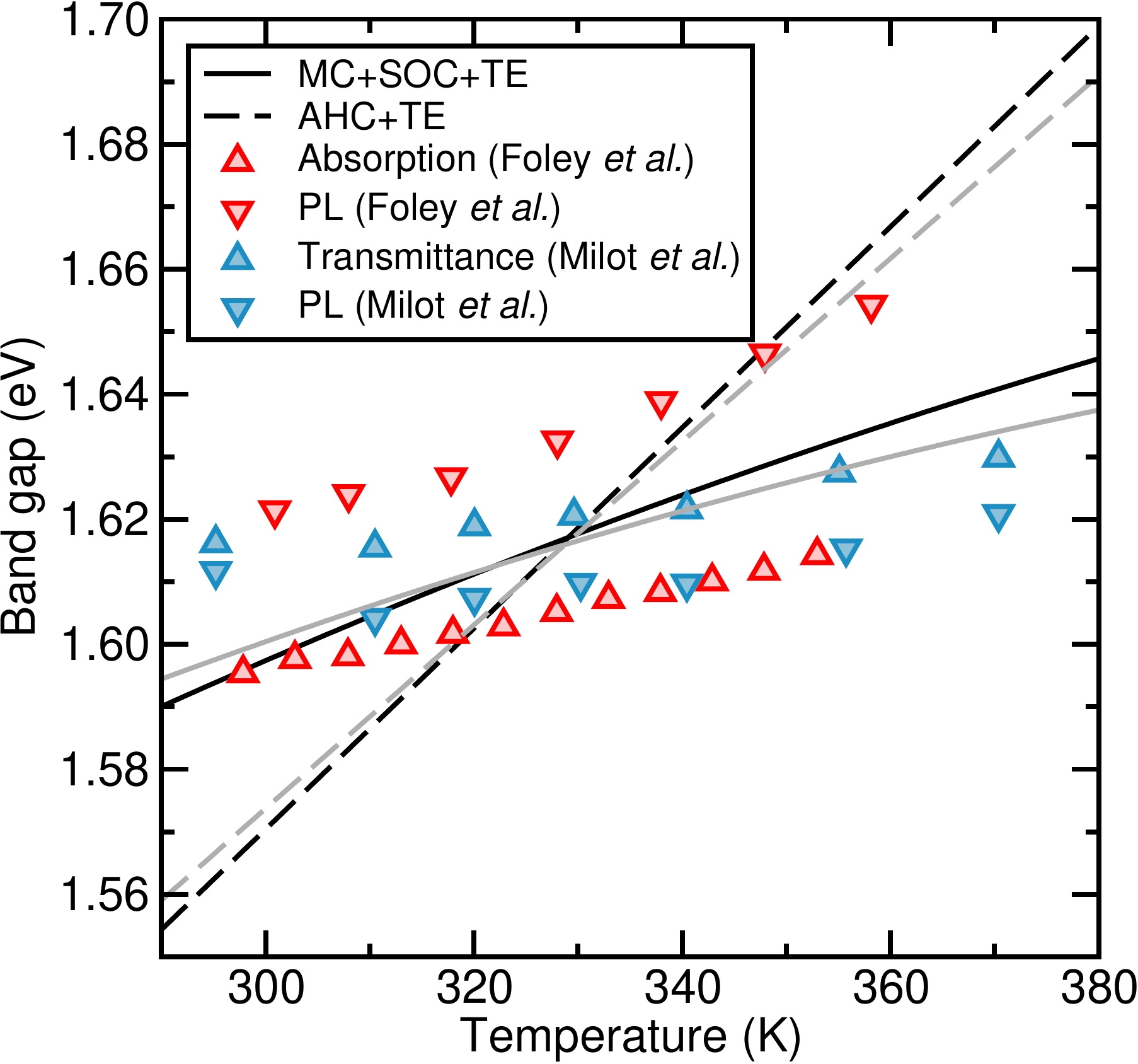}
\caption{Temperature dependence of the band gap of MAPbI$_3$ from
  Foley {\it et al.}~\cite{Foley_perovskites_2015} and from Milot {\it
    et al.}~\cite{Milot_2015_temperature_mapbi3} using various
  techniques (red and blue triangles respectively), compared to
  the theoretical prediction using the MC approach (black solid line)
  and the AHC theory (dashed black line). Both theoretical lines include
  the effects of thermal expansion, but only the MC line includes
  SOC. The results neglecting thermal expansion are shown with gray lines.} 
\label{fig:experiment_gap}
\end{figure}

Figure~\ref{fig:experiment_gap} shows the temperature dependence of
the MAPbI$_3$ band gap evaluated using the MC approach in conjunction
with PBE + SOC in the temperature range from $290$ to $380$~K,
relevant for solar cell applications. Thermal expansion effects on the
results are also presented in the figure.  For comparison, we show the
results obtained from two experimental studies measured using optical
absorbance or transmittance and photoluminescence
spectroscopy~\cite{Foley_perovskites_2015,Milot_2015_temperature_mapbi3}.
We only show the temperature range 290--380~K common to the two
studies, and we do not distinguish between the tetragonal and cubic
phases as argued before. The spread observed in the experimental data
from the two measurement techniques is likely due to the presence of
trap states below the band edges, but in both studies the band gap
variations with temperature are less sensitive to experimental
uncertainty and range between $30$--$40$~meV.  To avoid any
complications arising from well-known errors of DFT in computing
absolute band gaps, and to focus only on the band gap change with
temperature, we shift the calculated temperature dependence such that
at $329$~K the value of the gap is $1.616$~eV, given by the average of
the experimental data.

The change in the experimental band gap of $30$--$40$~meV in the temperature range of
interest is in good agreement with the MC results that
predict a change of $40$~meV. Repeating the MC calculations using a
$2\times2\times2$ supercell gives a change of $30$~meV, also within the
experimentally observed range.
The change of the band gap originates from the VBM as the computed CBM 
correction is relatively flat in that temperature range, which is 
consistent with experiment~\cite{Foley_perovskites_2015}.
By contrast, the AHC theory predicts a change of $120$~meV in the band gap, severely overestimating the strength of
electron-phonon coupling.
These results are the first direct comparison of two leading first-principles methods to compute the
temperature-dependence of band gaps where higher-order terms dominate,
and for which an experimental measurement exists.

Although the AHC theory fails to provide quantitatively accurate
results for the strength of electron-phonon coupling in MAPbI$_3$, it
is still useful in investigating the microscopic origin of the
electron-phonon coupling strength, as each normal mode
$(\mathbf{q},\nu)$ appears independently in Eq.~(\ref{eq:ahc}).  Our
results show that low-energy crystal modes dominate the coupling over
the molecular high-energy modes. This is consistent with the fact that
the VBM and CBM are formed by states whose character is dominated by
Pb and I. 
It is interesting to note that the
low-energy modes in MAPbI$_3$ also drive the tetragonal to cubic phase
transition as temperature increases~\cite{WAS_2016}. These results are also suggestive that the
failure of the AHC theory is not due to  high energy phonon
modes associated with the organic molecule as these contribute the
least to the energy renormalization. We have further investigated this
point by calculating the temperature dependence of the band gap
of the inorganic CsPbI$_3$ perovskite, where a similar picture emerges
(see Supporting Information). 

Overall, our results show that temperature drives the changes in the absolute
value of the band gap of MAPbI$_3$, as well as in the shape of the band 
structure, from both electron-phonon coupling and thermal expansion. It has
also been shown in previous work how the molecular orientation of the organic
cation can modify these quantities~\cite{mosconi_md_mapbi3,sanvito_bs_mapbi3},
and it would be interesting in future work to explore the combination of both
effects.


{\it Conclusions.} We have determined the temperature dependence of the band gap 
and band edges of the cubic phase of MAPbI$_3$ using first-principles methods. 
Our results show that the MAPbI$_3$ band gap opens by $40$~meV as temperature 
increases from $290$~K to $380$~K
due to thermal expansion and electron-phonon coupling, which is in
excellent agreement with the experimental value of $30$-$40$~meV. 
Furthermore, we demonstrate that the calculation of electron-phonon coupling is
particularly intricate, as we expose the limitations of the commonly used 
Allen-Heine-Cardona theory, arising from the dominant contribution in the
electron-phonon coupling of high-order terms. We have also found that
the use of the HSE hybrid functional has little effect, but the
presence of Pb atoms means that the spin-orbit interaction
substantially modifies the strength of electron-phonon coupling. 

Overall, our Monte Carlo approach leads to good agreement between theory and 
experiment, and unambiguously confirms for the first time the importance of
higher-order terms in the electron-phonon coupling. These results call for a 
revision of previous calculations of electron-phonon coupling in these materials, 
used for example to study their transport properties. We also hope that our 
findings will foster the design of new perovskite materials and the development of
improved first-principles methods to efficiently tackle these challenging materials.

\acknowledgements
W.A.S. thanks Joshua J. Choi, Christian Carbogno, and Benjamin
Folley for many useful discussions, and Laura Herz for 
sending the data used in Fig.~\ref{fig:experiment_gap}. 
We also thank Mireia Crispin-Ortuzar for help in designing the 
graphical abstract.
W.A.S. is supported in part by a start-up fund from the Department of Mechanical 
Engineering and Materials Science at the University of
Pittsburgh. B.M.~thanks Robinson College, Cambridge, and the Cambridge Philosophical Society for a Henslow Research Fellowship. Calculations were performed in part at the University of Pittsburgh Center for Simulation and Modeling. This work used the Extreme Science and Engineering Discovery Environment (XSEDE), which is supported by National Science Foundation grant number OCI-1053575.

 \bibliography{mapbi3}

\pagebreak
\widetext
\begin{center}
\textbf{\large Supporting Information for ``Temperature Dependence of the Energy Levels of Methylammonium Lead Iodide Perovskite from First Principles''}
\end{center}
\setcounter{equation}{0}
\setcounter{figure}{0}
\setcounter{table}{0}
\setcounter{page}{1}
\makeatletter
\renewcommand{\theequation}{S\arabic{equation}}
\renewcommand{\thefigure}{S\arabic{figure}}
\renewcommand{\bibnumfmt}[1]{[S#1]}
\renewcommand{\citenumfont}[1]{S#1}

\section{Computational details} \label{sec:computational_details}

Most calculations reported in the main text have been performed using the plane-wave DFT code {\sc vasp}~\cite{vasp1,vasp2,vasp3,vasp4}. We use the PBE~\cite{PBE} and HSE~\cite{hse06_functional,hse06_functional_erratum} functionals as described in the text, together with the projector augmented-wave method~\cite{paw_original,paw_us_relation}. The energy cut-off used is $500$~eV, and the electronic Brillouin zone (BZ) is sampled using a $6\times6\times6$ grid for the primitive cell, and commensurate grids for the supercells. Spin-orbit coupling (SOC) is included as described in the text.

The vibrational harmonic calculations have been done using the finite
displacement method in conjunction with nondiagonal
supercells~\cite{non_diagonal}. The electron-phonon coupling
calculations using the FD approach have also been performed using the
finite displacement approach and nondiagonal supercells. The
non-adiabatic Allen-Heine-Cardona calculations based on DFPT have been
performed with the {\sc abinit} software~\cite{Gonze2016} using a $6\times6\times6$ $\boldsymbol{\Gamma}$-centered $\mathbf{k}$-point grid with a plane-wave energy cutoff of $680$~eV. Norm-conserving pseudopotential with the PBE functional have been used. The vibrational BZ was sampled using grids of sizes up to $20\times20\times20$.

The cubic cell that we have used in our calculations exhibits some imaginary phonon frequencies, reflecting the dynamical instability of this structure, and driving it towards the lower-temperature tetragonal and orthorhombic structures. In the calculations reported in the main manuscript, we have set the amplitudes of the vibrational modes corresponding to imaginary frequencies to zero, to be able to work with the cubic structure. We have performed additional calculations for the $2\times2\times2$ supercell in which we have allowed the imaginary modes to contribute to electron-phonon coupling by describing them using Gaussian vibrational wave functions of amplitude given by the absolute value of their (imaginary) frequencies, in the spirit of the self-consistent harmonic approximation. Our calculations show that the temperature dependence does not change within error bars whether the unstable modes are allowed to contribute or not.

\section{Expansion order in the electron-phonon interaction}

The electronic state average over atomic vibrations
\begin{equation}
\epsilon_{\mathbf{k}n}(T)=\frac{1}{\mathcal{Z}}\sum_{\mathbf{s}}\langle\chi_{\mathbf{s}}|\epsilon_{\mathbf{k}n}|\chi_{\mathbf{s}}\rangle e^{-E_{\mathbf{s}}/k_{\mathrm{B}}T}, \label{eq:epsilon_temperature_sup}\end{equation}
is usually performed using a {\it quadratic approximation} to the dependence of $\epsilon_{\mathbf{k}n}$ on the atomic configuration. In terms of normal modes of vibration $u_{\mathbf{q}\nu}$ characterized by a wave vector $\mathbf{q}$ with branch $\nu$, the value of the electronic state at a general atomic configuration $\mathbf{u}=\{u_{\mathbf{q}\nu}\}$ can be expanded around its equilibrium position as
\begin{equation}
\epsilon_{\mathbf{k}n}(\mathbf{u})=\epsilon_{\mathbf{k}n}(\mathbf{0})+
\sum_{\mathbf{q},\nu}a^{(1)}_{\mathbf{q}\nu}u_{\mathbf{q}\nu}+\sum_{\mathbf{q},\nu,\mathbf{q}',\nu'}a^{(2)}_{\mathbf{q}\nu;\mathbf{q}'\nu'}u_{\mathbf{q}\nu}^*u_{\mathbf{q}'\nu'}+\cdots, \label{eq:quadratic_expansion}
\end{equation}
where $\{a^{(1)}_{\mathbf{q}\nu},a^{(2)}_{\mathbf{q}\nu;\mathbf{q}'\nu'},\ldots\}$ are the electron-phonon coupling constants. Truncating this expansion at second order and substituting into Eq.~(\ref{eq:epsilon_temperature_sup}) leads to
\begin{equation}
\Delta \epsilon_{\mathbf{k}n}(T) \triangleq \epsilon_{\mathbf{k}n}(T) -\epsilon_{\mathbf{k}n}(\mathbf{0}) 
= \frac{1}{N_{\mathbf{q}}}\sum_{\mathbf{q},\nu}\frac{a^{(2)}_{\mathbf{q}\nu;\mathbf{q}\nu}}{\omega_{\mathbf{q}\nu}}\left[\frac{1}{2}+n_{\mathrm{B}}(\omega_{\mathbf{q}\nu},T)\right], \label{eq:ahc_sup}
\end{equation}
where $\Delta \epsilon_{\mathbf{k}n}(T)$ is the phonon renormalization of the electronic state at temperature $T$, $N_{\mathbf{q}}$ is the number of points in the vibrational BZ, $\omega_{\mathbf{q}\nu}$ is the harmonic frequencies of mode $(\mathbf{q},\nu)$, and $n_{\mathrm{B}}$ is a Bose-Einstein factor. The expression in Eq.~(\ref{eq:ahc_sup}) is computationally convenient, as the coupling constants of interest only depend on individual modes, and can be efficiently calculated using density functional perturbation theory (DFPT)~\cite{Gonze1997a} or finite displacements~\cite{phonon_finite_displacement}.

To obtain a DFPT formulation of Eq.~(\ref{eq:ahc_sup}), the second-order coupling constants are expressed as~\cite{Ponce2014a}
\begin{equation}
a^{(2)}_{\mathbf{q}\nu;\mathbf{q}\nu} = \frac{1}{2}\Big[ \mathbf{F}_{\mathbf{q}\nu} u_{\mathbf{q}\nu}^* u_{\mathbf{q}\nu} +  \mathbf{D}_{\mathbf{q}\nu} u_{\mathbf{q}\nu}^* u_{\mathbf{q}\nu} \Big],
\end{equation}
where $\mathbf{F}_{\mathbf{q}\nu}=\frac{1}{2}[\langle \psi_{\mathbf{k,q},n}^{(1)} | \hat{H}_{\mathbf{k+q,k}}^{(1)} | \psi_{\mathbf{k}n}^{(0)}\rangle + (c.c.)]$ is the Fan term of the periodic part of the electronic wavefunction $|\psi^{(0)}\rangle$ with Hamiltonian $\hat{H}$, $\hat{H}^{(1)}$ is the first-order change of the Hamiltonian due to a lattice distortion arising from nuclear vibrations, and $|\psi^{(1)}\rangle$ is the corresponding first-order change in the wavefunction. $\mathbf{D}_{\mathbf{q}\nu} = \langle \psi_{\mathbf{k}n}^{(0)} | \hat{H}_{\mathbf{k-q,k+q}}^{(2)} | \psi_{\mathbf{k}n}^{(0)}  \rangle$ is the Debye-Waller (DW) term arising from the second-order change of the Hamiltonian due to nuclear vibrations. Using the rigid-ion approximation to recast the DW term in terms of first-order matrix elements that are directly accessible through routine DFPT calculations, one obtains the so-called Allen-Heine-Cardona theory~\cite{ahc_1976,Allen1981}.
Part of the DW term is neglected in the process, hence the approximation. The part of the DW term neglected by the rigid-ion approximation has been found to be small in crystals~\cite{Ponce2014a} but crucial for molecules~\cite{gonze_2011}. Thanks to this approximation, very fine BZ grids can be used to fully converge the results~\cite{ponce_2015}. In the main manuscript, this approximation is referred to as {\it Allen-Heine-Cardona theory} (AHC).

To obtain the finite displacement formulation of Eq.~(\ref{eq:ahc_sup}), the second-order coupling constants are expressed as
\begin{equation}
a^{(2)}_{\mathbf{q}\nu;\mathbf{q}\nu}=\frac{\epsilon_{\mathbf{k}n}(u_{\mathbf{q}\nu})+\epsilon_{\mathbf{k}n}(-u_{\mathbf{q}\nu})}{2u^2_{\mathbf{q}\nu}},
\end{equation}
and explicit atomic displacements $\pm u_{\mathbf{q}\nu}$ are used to evalute the changes in the electronic eigenenergies. The finite displacement formulation does not rely on the rigid-ion approximation, but the vibrational BZ can only be sampled by explicitly constructing supercells of the primitive cell. Until recently, the fine BZ grids needed to converge the evaluation of electron-phonon coupling in this context made the calculations using the finite displacement method prohibitive, due to the large supercell sizes required. This situation has improved significantly thanks to the introduction of nondiagonal supercells, that provide access to very fine BZ grids using moderate supercells sizes, and approaching the levels of convergence available using DFPT~\cite{non_diagonal}. In the main manuscript, this approximation is referred to as {\it finite displacements} (FD).

Despite the successes of the quadratic theory (AHC or FD) to describe the temperature dependence of the band structures of a range of materials, recent calculations question the validity of the AHC theory in helium at terapascal pressures~\cite{helium}, molecular crystals at ambient conditions~\cite{molec_crystals_elph}, in the perovskite CsSnI$_3$~\cite{patrick_cssni3_2015}, and even (albeit moderately) in crystals like diamond~\cite{Antonius2015} because higher-order terms in the expansion of Eq.~(\ref{eq:quadratic_expansion}) make significant contributions. These terms can be included by evaluating Eq.~(\ref{eq:epsilon_temperature_sup}) directly using Monte Carlo integration
\begin{equation}
\epsilon_{\mathbf{k}n}(T)\simeq\frac{1}{M}\sum_{i=1}^{M}\epsilon_{\mathbf{k}n}(\mathbf{u}_i), \label{eq:mc}
\end{equation}
where $M$ is the number of sampling points, distributed according to the vibrational density. The use of Monte Carlo integration leads to statistical error bars in the Monte Carlo estimates of the integrals, which are given by
\begin{equation}
\delta\epsilon_{\mathbf{k}n}(T)\simeq\left[\frac{1}{M(M-1)}\sum_{i=1}^M\left(\epsilon_{\mathbf{k}n}(\mathbf{u}_i)-\frac{1}{M}\sum_{j=1}^M\epsilon_{\mathbf{k}n}(\mathbf{u}_j)\right)^2\right]^{1/2}.
\end{equation}
Error bars are included in all results obtained using Monte Carlo integration, and if they cannot be seen in a given plot, that means that their size is smaller than the size of the point shown.

We emphasize here that the vibrational wave function is treated within the harmonic approximation, even when the higher order terms are included in the description of the coupling of vibrations to electronic eigenstates. Within the harmonic approximation, the vibrational density at temperature $T$ is given by a product of Gaussian functions over the normal modes $\prod_{\mathbf{q},\nu}(2\pi\sigma^2_{\mathbf{q}\nu})^{-1/2}\exp(-\frac{u_{\mathbf{q}\nu}^2}{2\sigma_{\mathbf{q}\nu}^2})$ of amplitude
\begin{equation}
\sigma^2_{\mathbf{q}\nu}(T)=\frac{1}{\omega_{\mathbf{q}\nu}}\coth\left(\frac{\omega_{\mathbf{q}\nu}}{k_{\mathrm{B}}T}\right),
\end{equation}
where $k_{\mathrm{B}}$ is Boltzmann's constant. While this approach fully accounts for all higher-order terms in electron-phonon coupling, the presence of cross-terms between different modes means that it cannot be used with DFPT or nondiagonal supercells, and therefore the computational advantages that these methods provide cannot be exploited. In the main manuscript, this approach is refereed to as the {\it Monte Carlo approach} (MC).

\begin{figure} \centering
\includegraphics[scale=0.43]{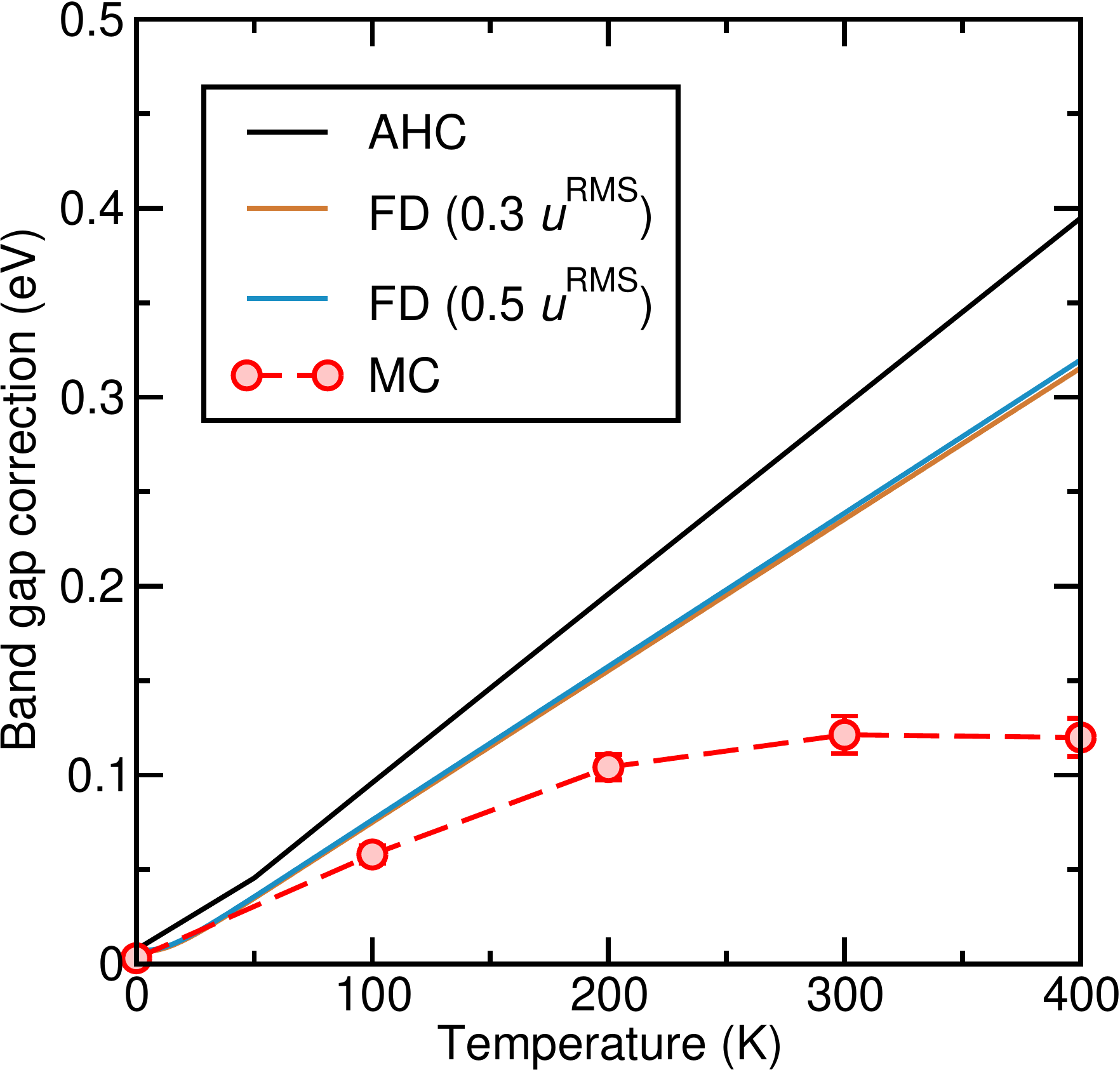}
\caption{Temperature dependence of the band gap of MAPbI$_3$ evaluated using the AHC theory within DFPT (black lines), the quadratic approximation within a FD approach (blue and orange lines), and the MC sampling approach (red circles). For the FD approach, two sets of calculations have been performed, with finite displacement amplitudes given by $0.3u^{\mathrm{RMS}}$ and $0.5u^{\mathrm{RMS}}$ where $u^{\mathrm{RMS}}=\sqrt{\langle u^2\rangle}$. The calculations have been performed by sampling a $2\times2\times2$ vibrational Brillouin zone grid, and without spin-orbit coupling. The statistical error bars in the MC results are included, but for the low temperature data points their size is smaller than the symbol size.}
\label{fig:rigid_ion}
\end{figure}

In Fig.~\ref{fig:rigid_ion} we compare the temperature dependence of the band gap of MAPbI$_3$ using the three different methods described above. The two approaches based on the quadratic expansion of Eq.~(\ref{eq:quadratic_expansion}), AHC and FD, show a significant increase in the value of the band gap with temperature, reaching $0.3$-$0.4$~eV at $400$~K. The disagreement between the two quadratic methods approaches $70$~meV at $400$~K. The origin of this discrepancy could be attributed to various factors. First, the AHC calculations have been performed with the {\sc abinit} package~\cite{Gonze2016}, while the FD calculations have been performed with the {\sc vasp} package~\cite{vasp1,vasp2,vasp3,vasp4}. Different pseudopotentials and slightly different atomic coordinates have been used (although the volumes were the same).

In Fig.~\ref{fig:rigid_ion} we compare the temperature dependence of the band gap of MAPbI$_3$ using the three different methods described above. The two approaches based on the quadratic expansion of Eq.~(\ref{eq:quadratic_expansion}), AHC and FD, show a significant increase in the value of the band gap with temperature, reaching $0.3$-$0.4$~eV at $400$~K. The disagreement between the two quadratic methods approaches $70$~meV at $400$~K. The origin of this discrepancy could be attributed to various factors. First, the AHC calculations have been performed with the {\sc abinit} package~\cite{Gonze2016}, while the FD calculations have been performed with the {\sc vasp} package~\cite{vasp1,vasp2,vasp3,vasp4}. Different pseudopotentials and slightly different atomic coordinates have been used (although the volumes were the same).
Given the above-mentioned differences, the agreement between the two methods is quite satisfactory. This leads us to believe that the rigid-ion approximation involved in the AHC approach seems to hold for molecular crystals. A more throughout investigation would be required to confirm this.

In any case, Fig.~\ref{fig:rigid_ion} clearly shows that the quadratic methods are not appropriate for the description of electron-phonon coupling in MAPbI$_3$, as they significantly overestimate the strength of the electron-phonon coupling. The higher-order terms included in the MC approach lead to a different temperature dependence of the band gap, and as shown in the main manuscript, agreement with experiment is only obtained if these higher-order terms are included.

\section{Electronic structure method}

In the main manuscript, we have established that the inclusion of SOC is necessary for an accurate description of the strength of the electron-phonon coupling in MAPbI$_3$, but that nonlocal electronic correlations are not important. Here, we provide further details of those calculations.

In evaluating Eq.~(\ref{eq:epsilon_temperature_sup}), both electronic states and phonons are usually treated within semilocal DFT. However, semilocal DFT is known to severely underestimate band gaps in semiconductors and insulators~\cite{band_gap_problem_i,band_gap_problem_ii}. In MAPbI$_3$, the situation is somewhat complex. The experimental band gap is $1.6$~eV~\cite{experiment_mapbi3_gap}, and semilocal DFT predicts a band gap of about $1.6$~eV~\cite{mapbi3_dft_nsoc_gap}, which appears to be in good agreement with experiment. Nevertheless, the presence of Pb atoms means that SOC is strong in MAPbI$_3$, and its inclusion in the calculations reduces the band gap by around $1.0$~eV~\cite{mapbi3_dft_soc_gap}, so that indeed semilocal DFT underestimates the band gap. Performing calculations using hybrid functionals including SOC brings the gap closer to the experimental value, and $GW$+SOC calculations lead to band gaps in very good agreement with experiment~\cite{mapbi3_gw_soc_gap,mapbi3_scgw_soc_gap,mapbi3_gw_soc_gap_giustino}. It had been assumed for some time that, while electronic states were poorly reproduced by semilocal DFT, electron-phonon coupling was well-described by this level of theory, as only {\it changes} in band gaps need to be calculated. However, it has recently been shown that in some materials semilocal DFT severely underestimates the strength of electron-phonon coupling~\cite{antonius_gw_elph,gw_thermal_lines}.

\begin{figure} \centering
\includegraphics[scale=0.43]{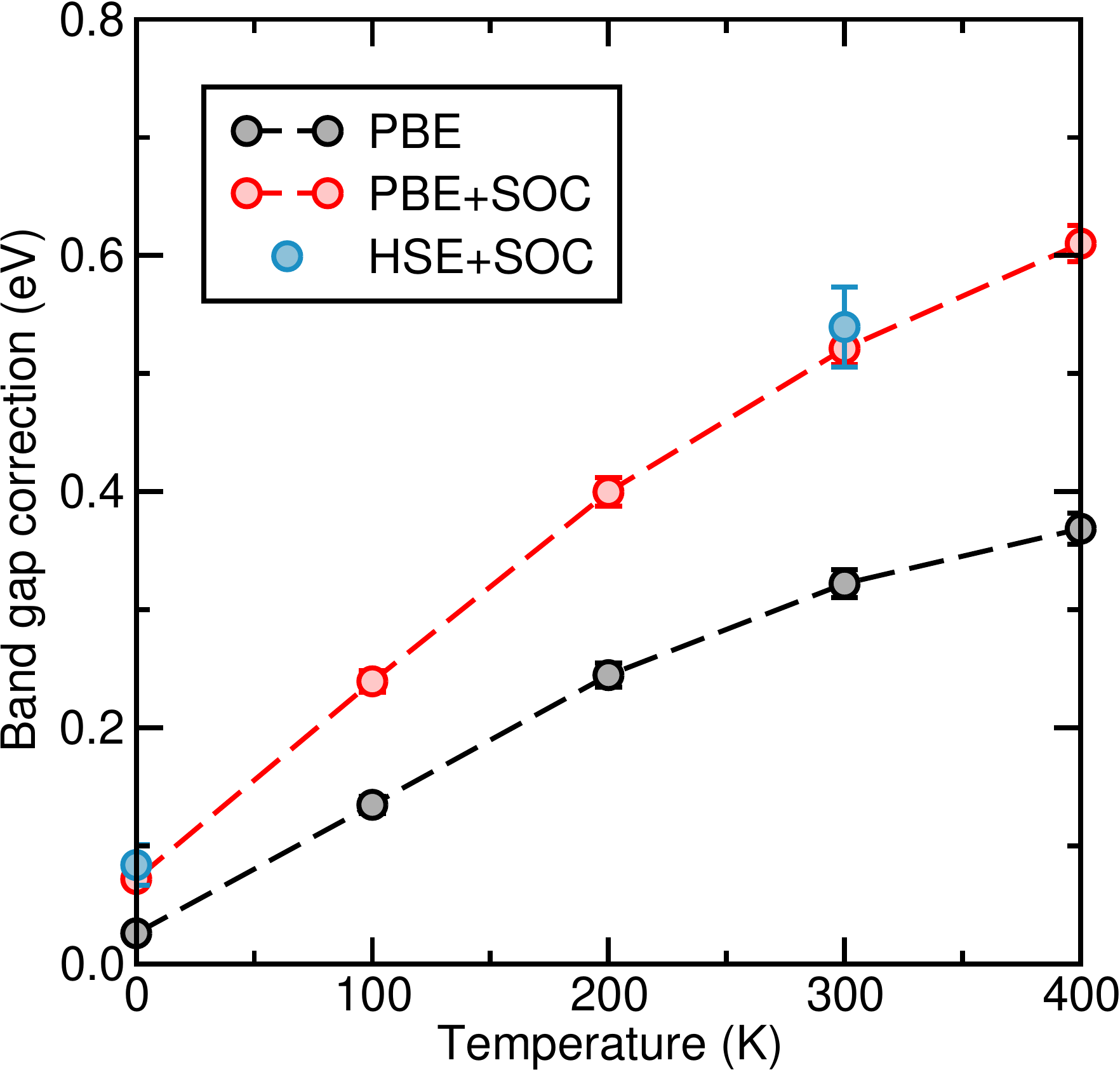}
\caption{Temperature dependence of the band gap of MAPbI$_3$ evaluated using PBE (black circles), PBE+SOC (red circles), and HSE+SOC (blue circles), and the MC method. Only the electron-phonon coupling constants for $\boldsymbol{\Gamma}$-point phonons are included. The statistical error bars are included in all data points, although they are not visible in some as their size is smaller than the symbol size.}
\label{fig:electronic_structure_method}
\end{figure}

Using MC sampling, we evaluate the temperature dependence of the band gap of MAPbI$_3$ using PBE, PBE+SOC, and HSE+SOC, by focusing again on the $\boldsymbol{\Gamma}$-point phonons. The results are shown in Fig.~\ref{fig:electronic_structure_method}. The inclusion of SOC significantly increases the strength of the electron-phonon coupling, such that at $500$~K, the band gap change is underestimated by $0.28$~eV if the SOC is neglected. The effects of SOC arise exclusively from the CBM, which is mostly of Pb character. For example, at $300$~K, the change in the VBM is $-0.24$~eV both without and with SOC, but the change in the CBM is $+0.08$~eV without SOC, and increases to $+0.28$~eV with SOC.

The use of the hybrid functional HSE~\cite{hse06_functional,hse06_functional_erratum} instead of PBE does not seem to make a significant difference. As a consequence, the final results in the main manuscript are obtained using PBE+SOC.

\section{Brillouin zone sampling}

Using the quadratic theory of Eq.~(\ref{eq:ahc_sup}), it has been shown that the strength of electron-phonon coupling converges slowly as a function of vibrational BZ grid size~\cite{ponce_2015,non_diagonal}. Unfortunately, the quadratic theory is not applicable to MAPbI$_3$ due to the importance of high-order terms, and that the methods developed to accurately sample the BZ cannot be used here. For the MC sampling method, supercells of the primitive cell need to be constructed, and their sizes are limited by the maximum number of atoms that can be realistically included in a DFT calculation. In our case, we find that we can reach system sizes of $3\times3\times3$ primitive cells, containing $324$ atoms.
\begin{figure}
\centering
\includegraphics[scale=0.43]{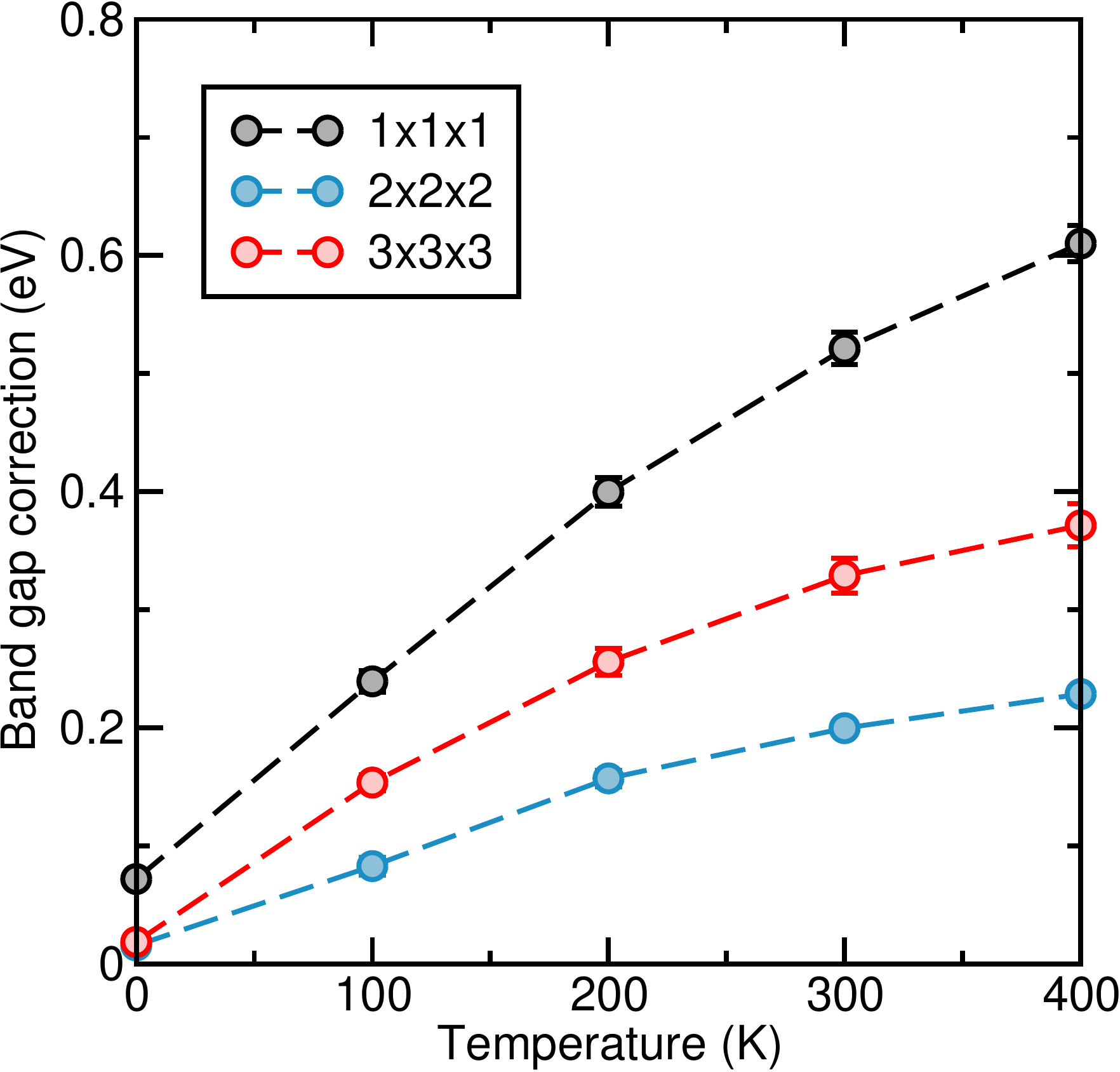}
\caption{Temperature dependence of the band gap of MAPbI$_3$ evaluated using supercells of sizes $1\times1\times1$ (black circles), $2\times2\times2$ (blue circles), and $3\times3\times3$ (red circles). The statistical error bars are included in all data points, although they are not visible in some as their size is smaller than the symbol size.}
\label{fig:sc}
\end{figure}

In Fig.~\ref{fig:sc} we show the temperature dependence of the band gap of MAPbI$_3$ for supercells of varying sizes, and it is clear that the band gap change is not converged for the largest supercells studied. Nonetheless, the changes arising from using supercells of varying size are smaller than either the changes induced by including higher-order terms or the SOC interaction. Furthermore, the change in the band gap at the relevant temperatures for solar cell applications, of above $300$~K, is similar between the $2\times2\times2$ and the $3\times3\times3$ supercells, and the difference is smaller than the experimental uncertainty in the available data. Therefore, our final results in the main manuscript are reported for the $3\times3\times3$ supercell.

\section{Results using the AHC theory and MC sampling}

\begin{figure} \centering
\includegraphics[scale=0.43]{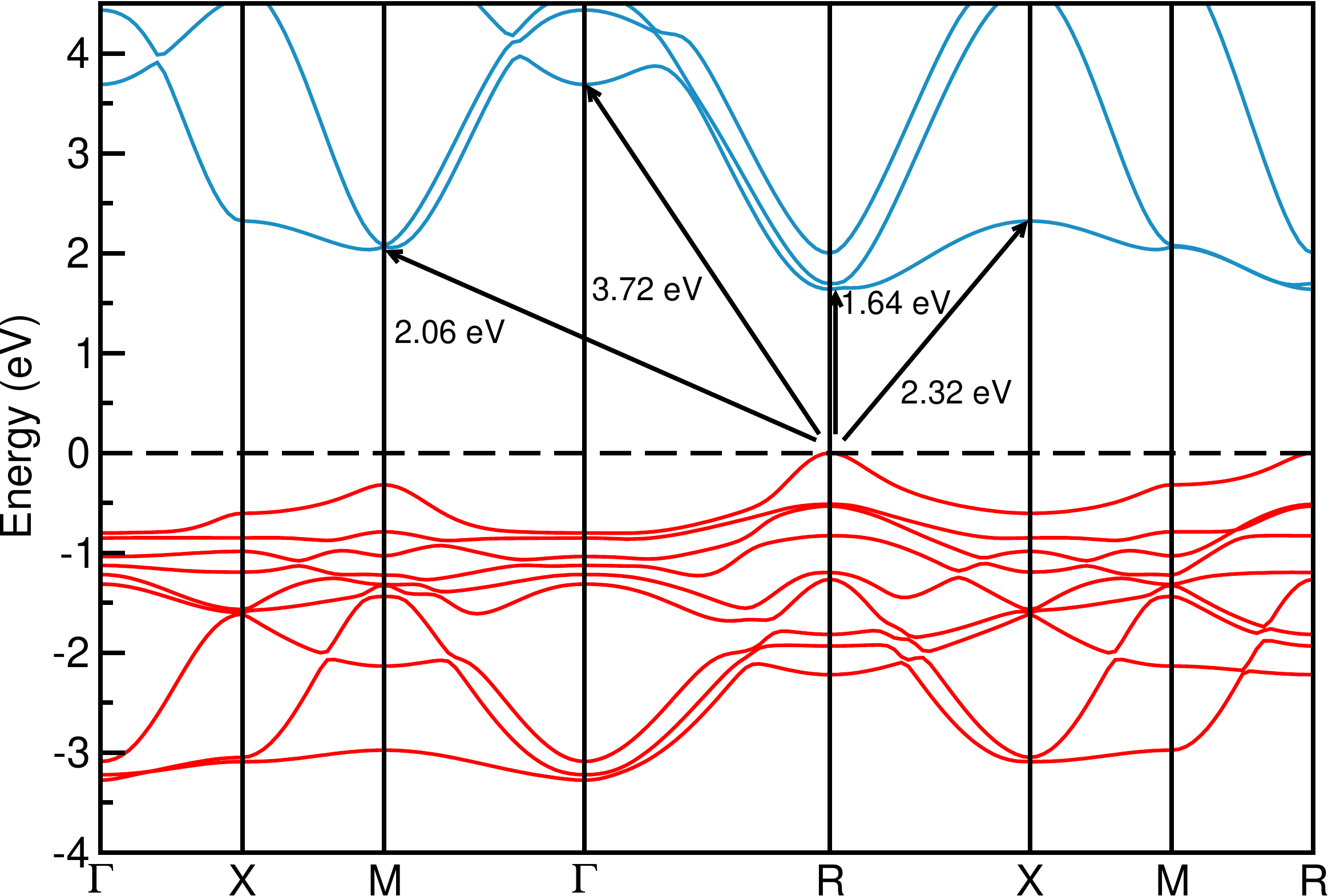}
\caption{Static lattice band structure of MAPbI$_3$ calculated using the PBE functional without including SOC. The dashed line indicates the position of the Fermi level.}
\label{fig:static_band_structure}
\end{figure}

\begin{figure} \centering
\includegraphics[scale=0.40]{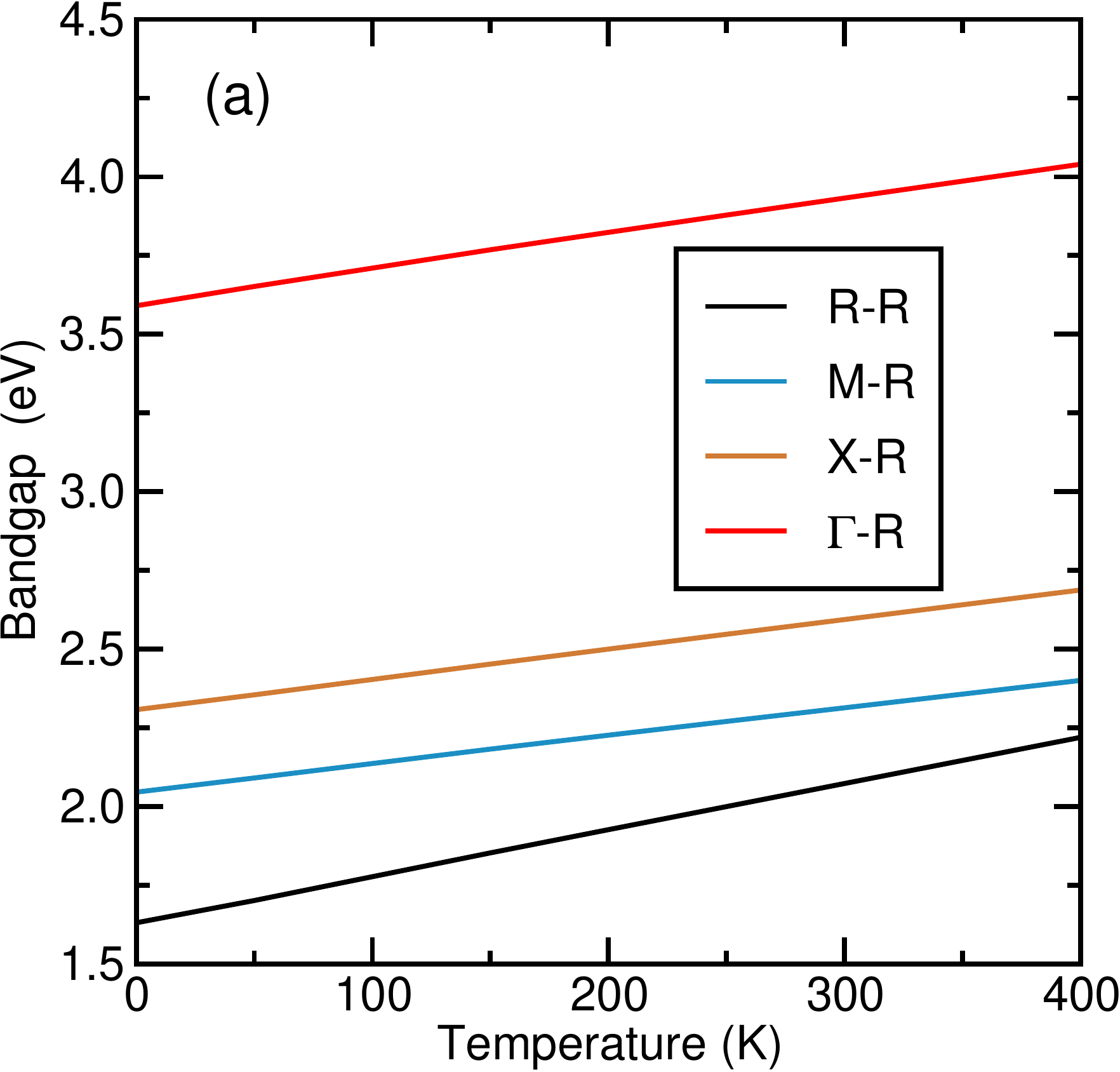}
\includegraphics[scale=0.40]{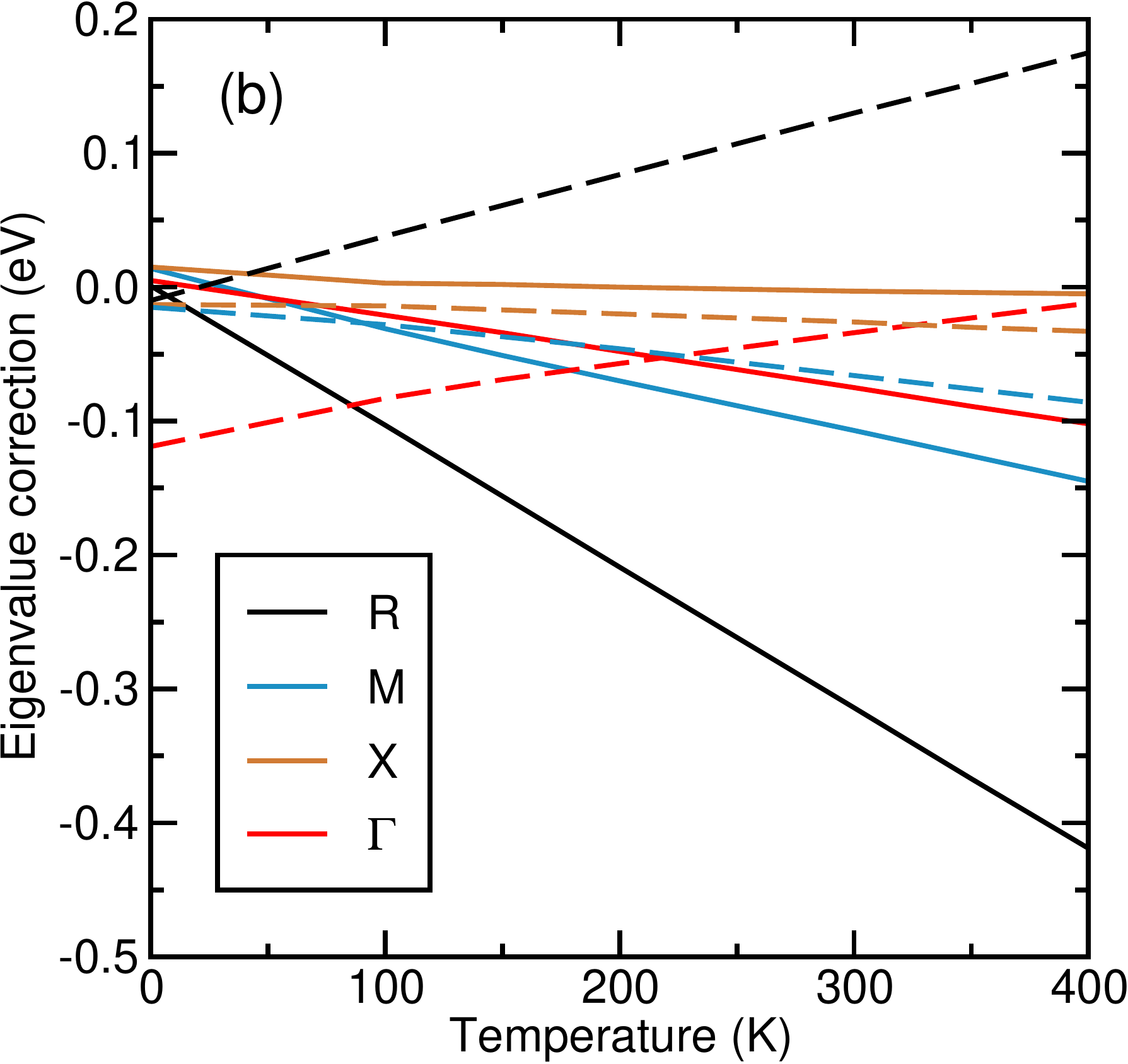}
\caption{(a) Temperature dependence of band gaps of MAPbI$_3$ from the VBM at the $\mathbf{R}$ point to the conduction band of several high-symmetry points in the BZ. (b) Temperature dependence of the individual eigenvalues at the valence (solid lines) and conduction (dashed lines) band at several high-symmetry points of the BZ. All calculations have been performed using the PBE functional and the AHC theory. The SOC and thermal expansion are not included.}
\label{fig:abinit_gaps}
\end{figure}

In this section we describe several results obtained using the AHC theory and MC sampling. In Fig.~\ref{fig:static_band_structure} we show the static lattice band structure of MAPbI$_3$ calculated using the PBE functional without the inclusion of SOC. The values of several band gaps from the valence band maximum (VBM) at the $\mathbf{R}=(0.5,0.5,0.5)$ point to the conduction band of several high-symmetry points are indicated by the arrows. The conduction band minimum (CBM) is also located at the $\mathbf{R}$ point. In Fig.~\ref{fig:abinit_gaps} we show the temperature dependence of these band gaps, calculated using the AHC theory with a $20\times20\times20$ BZ sampling. The different changes with temperature exhibited by the different gaps indicate that temperature does not only change the absolute value of the band gaps, but also changes the shape of the band structure.
Fig.~\ref{fig:abinit_gaps} also shows the temperature dependence of the individual eigenvalues at the valence and conduction bands at several high-symmetry points in the vibrational BZ. For the minimum band gap at $\mathbf{R}$ discussed in the main text, the VBM decreases with increasing temperature, and the CBM increases with increasing temperature.

\begin{figure} \centering
\includegraphics[scale=0.40]{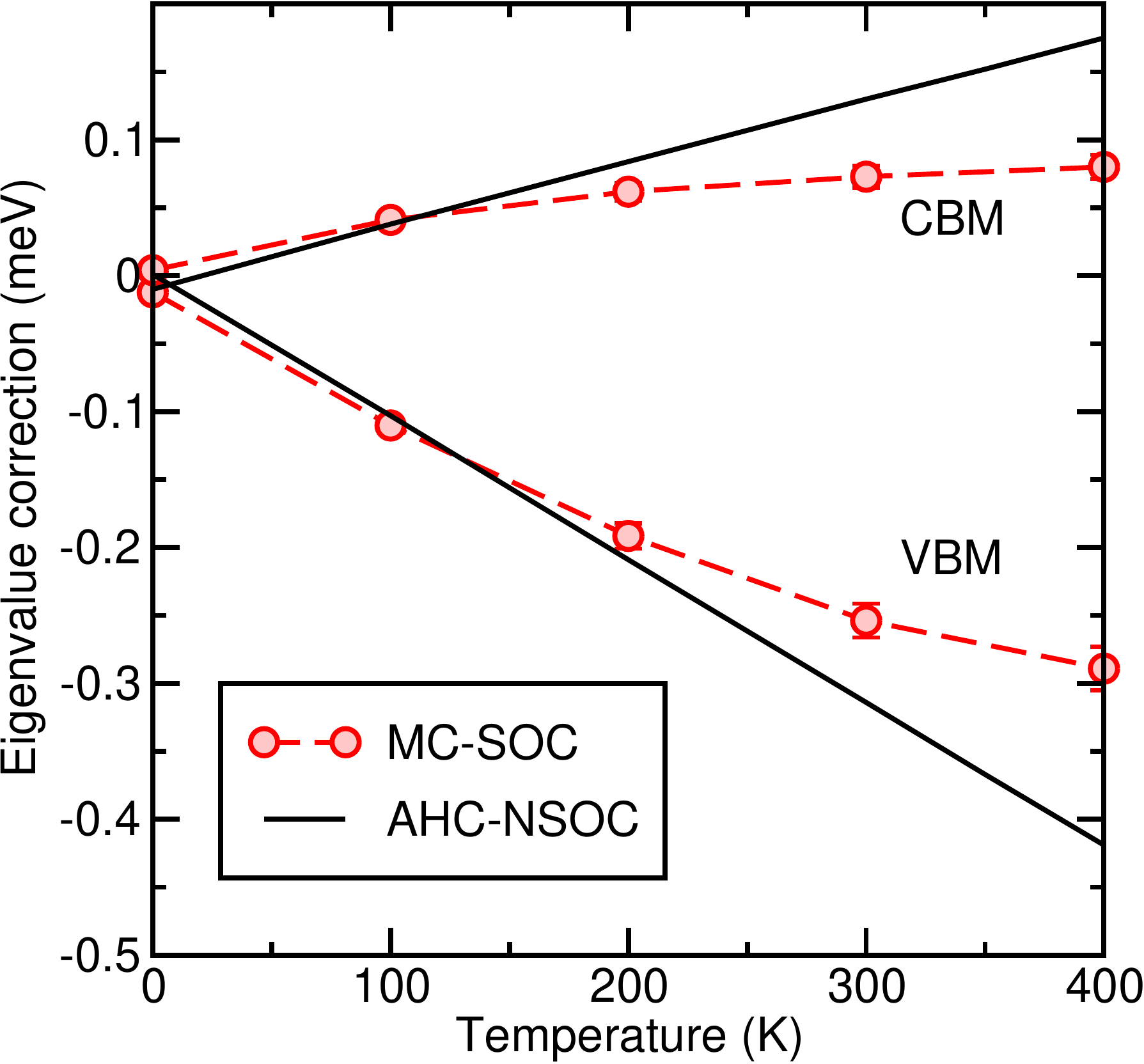}
\caption{Temperature dependence of the individual eigenvalues at the valence (lower lines) and conduction (upper lines) band at the $\mathbf{R}$-point of the BZ. The calculations have been performed using the same numerical parameters as those in Fig.~3 of the main manuscript, but without the inclusion of thermal expansion.}
\label{fig:mc_vbm_cbm}
\end{figure}

In Fig.~\ref{fig:mc_vbm_cbm} we show the temperature dependence of the VBM and CBM at the $\mathbf{R}$ point using MC sampling with the PBE+SOC method, and a supercell of size $3\times3\times3$, numerical parameters corresponding to the final results presented in the main manuscript. For comparison, we also repeat the results arising from using the AHC theory. These calculations do not include the effects of thermal expansion, as a proper treatment of the average electrostatic potential in simulation cells of varying volumes would be required, but this is beyond the scope of the present work. The results in Fig.~\ref{fig:mc_vbm_cbm} show that the use of MC sampling corrects to a large extent the defficiencies of the AHC approach, by almost removing the temperature dependence of the CBM at temperatures above $290$~K, of experimental interest. The results in Ref.~\cite{Foley_perovskites_2015} suggest that the inclusion of thermal expansion would make the temperature dependence of the CBM negative, in agreement with the experimental observations.

\section{Thermal expansion}

We study thermal expansion within the quasiharmonic approximation. The Helmholtz free energy of a solid at temperature $T$ can be written as
\begin{equation}
\mathcal{F}(V,T)=\mathcal{U}(V,T)+E_{\mathrm{vib}}(V,T),
\end{equation}
where $V$ is the volume, $\mathcal{U}$ is the electronic energy, and $E_{\mathrm{vib}}$ represent the vibrational energy. For systems with a band gap, $\mathcal{U}(V,T)\approx\mathcal{U}(V)$. The equilibrium volume at temperature $T$ is determined by minimizing $\mathcal{F}(V,T)$, which we do by calculating the vibrational energy within the harmonic approximation at a range of volumes and then directly minimizing $\mathcal{F}(V,T)$.

The calculations have been performed using density functional theory
(DFT) together with the Tkatchenko-Scheffler van der Waals
scheme~\cite{ts_vdW}. Using this approach, we find that the lattice
constant at zero temperature of $6.37$~\AA\@ increases to $6.40$~\AA\@
at $300$~K, which is in good agreement with experiment.\cite{WAS_2016}

\begin{figure} \centering
\includegraphics[scale=0.43]{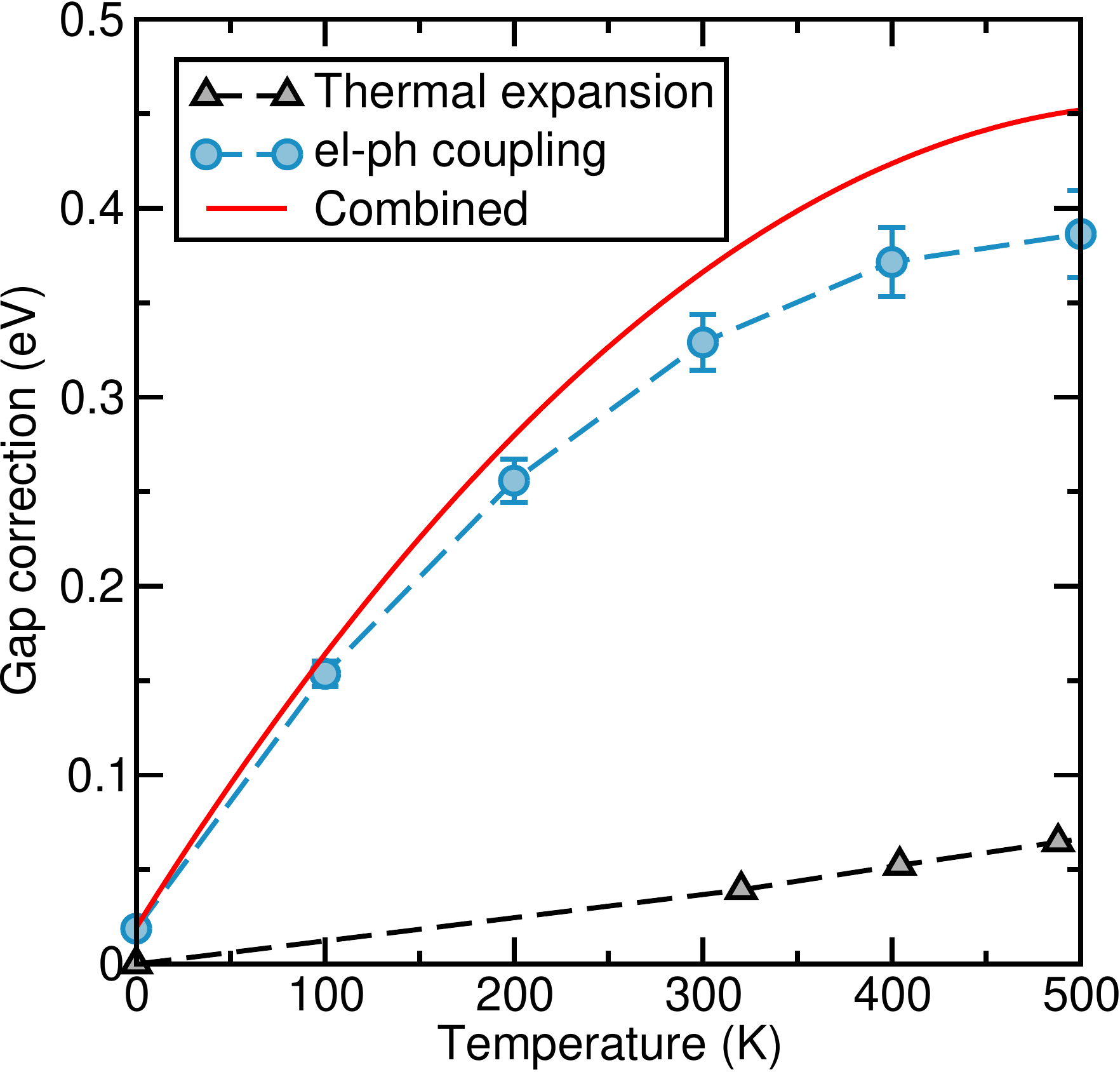}
\caption{Temperature dependence of the band gap of MAPbI$_3$ evaluated using thermal expansion only (black triangles), electron-phonon coupling only with MC and SOC on a $3\times3\times3$ supercell (blue circles), and both (red dashed-dotted line). The statistical error bars are included in all MC data points, although they are not visible in some as their size is smaller than the symbol size.}
\label{fig:thermal_expansion}
\end{figure}

The change in the band gap due to thermal expansion is shown in Fig.~\ref{fig:thermal_expansion}, where it is compared with the electron-phonon coupling induced change. Electron-phonon coupling makes the largest contribution, with a band gap opening of about $0.40$~eV at $500$~K, compared to $0.07$~eV for thermal expansion.

\section{\texorpdfstring{C\MakeLowercase{s}P\MakeLowercase{b}I$_3$}{CsPbI$_3$}}

\begin{figure} \centering
\includegraphics[scale=0.43]{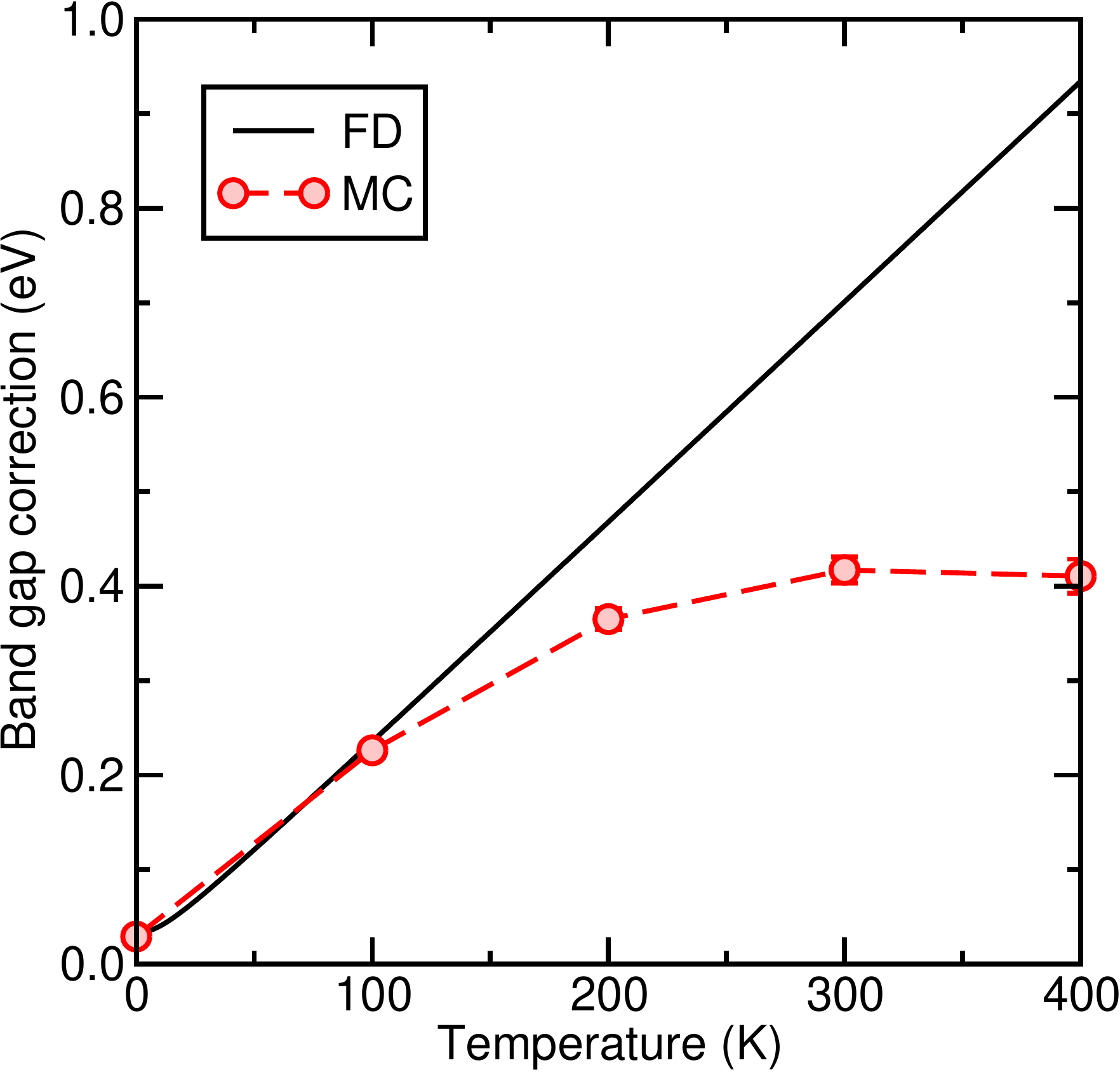}
\caption{Temperature dependence of the band gap of CsPbI$_3$ evaluated using the FD and MC methods. The calculations have been performed using the PBE functional, with a $2\times2\times2$ BZ grid, and without including SOC. Statistical error bars are included in the MC data points, but they are not visible for some data points as they are smaller than the symbols.}
\label{fig:cspbi3}
\end{figure}

In Fig.~\ref{fig:cspbi3} we show the temperature dependence of the band gap of CsPbI$_3$ evaluated using the quadratic approximation within the FD approach, and also using MC sampling. The calculations correspond to a sampling of the BZ with a grid of size $2\times2\times2$ and do not include SOC. As observed for MAPbI$_3$ in the main manuscript, high-order terms in the electron-phonon coupling are also necessary to accurately describe the temperature dependence of CsPbI$_3$.


\end{document}